# Observation of out-of-plane spin texture in a SrTiO$_3$ (111) two-dimensional electron gas

Pan He[1†], S. McKeown Walker[2†], Steven S.-L. Zhang[3], F. Y. Bruno[2], M. S. Bahramy[4,5], Jongmin Lee[1], Rajagopalan Ramaswamy[1], Kaiming Cai[1], Olle Heinonen[3], Giovanni Vignale[6], F. Baumberger[2,7], and Hyunsoo Yang[1]*

[1]*Department of Electrical and Computer Engineering, and NUSNNI, National University of Singapore, 117576, Singapore*

[2]*Department of Quantum Matter Physics, University of Geneva, 24 Quai Ernest-Ansermet, 1211 Genève 4, Switzerland*

[3] *Materials Science Division, Argonne National Laboratory, Lemont, Illinois 60439, USA*

[4]*Quantum-Phase Electronics Center, Department of Applied Physics, The University of Tokyo, Tokyo 113-8656, Japan.*

[5]*RIKEN Center for Emergent Matter Science (CEMS), Wako 351-0198, Japan. 8*

[6]*Department of Physics and Astronomy, University of Missouri, Columbia Missouri 65211, USA*

[7]*Swiss Light Source, Paul Scherrer Institut, CH-5232 Villigen PSI, Switzerland*

†These authors contributed equally to this work. *E-mail: eleyang@nus.edu.sg

**We explore the second order bilinear magnetoelectric resistance (BMER) effect in the *d*-electron-based two-dimensional electron gas (2DEG) at the SrTiO$_3$ (111) surface. We find an evidence of a spin-split band structure with the archetypal spin-momentum locking of the Rashba effect for the in-plane component. Under an out-of-plane magnetic field, we find a BMER signal that breaks the six-fold symmetry of the electronic dispersion, which is a fingerprint for the presence of a momentum dependent out-of-plane spin component. Relativistic electronic structure calculations reproduce this spin-texture and indicate that the out-of-plane component is a ubiquitous property of oxide 2DEGs arising from strong crystal field effects. We further show that the BMER response of the SrTiO$_3$ (111) 2DEG is tunable and unexpectedly large.**

In the presence of inversion-symmetry-breaking, Rashba-type spin-orbit coupling (SOC) lifts the spin degeneracy, which underlies a wide variety of fascinating phenomena [1]. Two-dimensional electron gases (2DEGs) formed at surfaces and interfaces have provided canonical examples of this and have been intensively studied because of their potential for spintronic applications [2-7]. In spite of only modest atomic SOC in $SrTiO_3$ (STO), 2DEGs based on STO (001) have recently emerged as strong candidates for oxide spintronics following the observation of highly efficient charge-spin interconversion in 2DEGs both at the STO surface [8] and $LaAlO_3/SrTiO_3$ (LAO/STO) interface [6,9,10]. High mobility 2DEGs have also been engineered at STO (111) surfaces and interfaces [11-14]. These studies were motivated by predictions for the existence of novel topological and multiferroic phases in (111) bilayers of $ABO_3$ cubic perovskites [15,16]. As shown in Fig. 1(a) and 1(b), the B site ions (Ti) of a bilayer resemble a honeycomb lattice, similar to that of graphene and topological insulators such as $Bi_2Se_3$.

Theoretical studies of the spin texture in STO (001) 2DEGs have revealed a picture of unconventional Rashba spin-splitting enhanced at the avoided crossings of subbands with orthogonal orbital character [17-21]. This picture supports experimental an evidence of SOC in LAO/STO 2DEGs, such as gate-tunable magnetotransport [22-24] and the unusual gate voltage dependence of spin-charge conversion efficiency [6]. In contrast, little is known about the spin texture of STO (111) 2DEGs or generally about the effect of surface orientation on spin structure.

Magnetoresistance that scales linearly with both the applied electric and magnetic fields has recently been discovered in a three-dimensional (3D) polar semiconductor [25] and a topological insulator [26]. This bilinear magnetoelectric resistance (BMER) causes a second order current density component $\mathbf{J}'(\mathbf{E}^2)$ in the presence of modest magnetic fields (**H**), and thus low [Fig. 1(e)] and high [Fig. 1(f)] conductance states for opposite electric fields (**E**) [25,26]. Such a non-reciprocal charge conductivity can occur in non-magnetic systems which have a spin-split Fermi surface with momentum-dependent spin textures [25-27]. In the presence of a magnetic field, which breaks time reversal symmetry, the Fermi surface of such a system can deform, permitting non-zero second order charge currents [25-27].

The nature of the Fermi surface deformation and thus the symmetry of the BMER signal with respect to the crystallographic axes depends on the zero-field spin texture and the direction of the magnetic field [26]. Taking advantage of this, BMER experiments have emerged as a sensitive probe of spin textures. This was demonstrated in $Bi_2Se_3$ where the modulations of the BMER signal induced by changing the directions of both magnetic and electric fields with respect to the crystallographic axes, were shown to be signatures of hexagonal warping in the topological surface state (TSS) [26]. In the 3D polar semiconductor BiTeX (X = I, Br), which hosts Rashba spin-split bulk bands [25], the BMER signal was shown to be a direct probe of intrinsic band structure parameters such as the effective mass and the Rashba SOC strength [25]. Note that the BMER has a different origin from the unidirectional spin Hall magnetoresistance found in ferromagnet/normal metal bilayers [28,29], which relies on a ferromagnetic material to provide spin-dependent asymmetric scatterings.

In this work, we report the observation of a BMER signal in the *d*-orbital 2DEG at the STO (111) surface [Fig. 1(a) and 1(b)], demonstrating spin-splitting. We show that our BMER measurements reveal a three-fold symmetric out-of-plane spin component that breaks the six-fold symmetry of the 2DEG subband dispersion, and an in-plane spin component locked perpendicularly to the momentum. By performing tight-binding supercell calculations based on the relativistic density functional theory of the STO bulk band structure, we find that this 3D spin texture is fully described by the effects of confinement of the STO $t_{2g}$ conduction band in the (111) plane. We also show that the BMER can be substantially tuned through oxygen vacancy doping, electrostatic gating and temperature variation. Our findings highlight the untapped potential of SrTiO$_3$ (111) based 2DEGs as a playground for spintronic applications.

Our BMER measurements were performed on the 2DEG formed by Ar$^+$-irradiation of the surface of commercial, insulating single crystals of STO (111) [8,30,31]. The 2DEG emerges due to quantum confinement of electrons by a band-bending potential near the surface, induced by oxygen vacancies created by the Ar$^+$-irradiation [13]. By tuning the irradiation time and power we can control the carrier density of the 2DEG [30,32,33]. The thickness and dimensionality of the electron gas are discussed in Supplemental Material [32], which includes Refs. [34-37]. We created 2DEGs with a Hall bar geometry as shown in Fig. 1(c) using standard photolithography techniques [32]. The experimental geometry is shown in Fig. 1(d). The angle $\varphi$ ($\theta$) of the in-(out-of-) plane magnetic field scans is defined with respect to the $x(z)$ axis. Multiple Hall bars with different orientations $\Theta$ were patterned on a single wafer, where $\Theta$ is the angle between the current channel ($x$ coordinate) and the [$\bar{1}$10] crystallographic axis of STO, as schematically shown in Fig. 1(b). The channel resistance as a function of temperature in Fig. 2(a) shows a metallic behavior similar to previous reports [8,30,31]. The Hall carrier density $n_H$ was extracted from the Hall data.

The linear current dependence of BMER implies a quadratic dependence of the current density on the electric field [32]. In order to disentangle the BMER signal form the current-independent resistivity, we apply an a.c. current $I_\omega = I\sin(\omega t)$ while measuring the longitudinal voltage of the first harmonic ($V_\omega$) and second harmonic ($V_{2\omega}$) simultaneously, and we extract the second harmonic resistance $R_{2\omega}$ from $V_{2\omega}$ [32]. Figure 2(b) shows the in-plane (*xy*) field-angle ($\varphi$) dependence of $R_{2\omega}$ for an STO (111) 2DEG with $n_H \approx 5.3 \times 10^{13}$ cm$^{-2}$. $R_{2\omega}$ shows a sinusoidal field-angle dependence with a period of 360° demonstrating that $R_{2\omega}$ changes sign when the magnetic field direction is reversed in contrast to the normal resistance $R_\omega$. The magnitude of $\Delta R_{2\omega}$ shows a linear dependence on the current [Fig. 2(c)] and magnetic field [Fig. 2(d)], confirming that it is a nonlinear effect induced by the magnetic field [32]. The observed BMER thus demonstrates a nonreciprocal *d*-electron magnetotransport in the STO (111) 2DEG. Since the contribution of thermoelectric effects to the BMER signal can be ruled out in our STO (111) sample [32], the BMER provides clear evidence for the spin-split nature of the Fermi surface. To investigate the 3D spin texture, we measure the BMER by applying currents along different crystallographic directions with in-plane and out-of-plane magnetic fields.

Figures 3(a-c) show the dependence of $R_{2\omega}$ on the magnetic field angles $\varphi$ and $\theta$ in the *xy*, *zy* and *zx* planes [see Fig. 1(d)] for Hall bar angles $\Theta = 0$, 30, and 60° for an STO (111) 2DEG with $n_H \approx 1.3 \times 10^{14}$ cm$^{-2}$. The data have been normalized to the maxima of each *xy* scan. The $R_{2\omega}$ signal for the *xy* scans (red) in Fig 3(a-c) has the same field-angle dependence regardless of the Hall bar orientation $\Theta$. The observation that $R_{2\omega}$ approaches zero when the magnetic field is parallel ($\varphi = 0°$) or antiparallel ($\varphi = 180°$) to the current direction, and has a peak value ($\Delta R_{2\omega}$) when the current and the magnetic field are perpendicular ($\varphi = 90°$ or $270°$), indicates that there is an in-plane spin component locked perpendicularly to the in-plane momentum. There is also a clear $R_{2\omega}$ signal for the *zx* scans (blue) with maximum intensities at $\theta = 0$ and $180°$ in Fig. 3(a) and 3(c), when **H** is perpendicular to the sample plane, which strongly suggests the presence of an out-of-plane spin component.

The $R_{2\omega}$ signal of the out-of-plane field scans also exhibits a striking dependence on the current direction. Specifically, $R_{2\omega}$ of the *zx* scan has opposite signs for $\Theta = 0°$ and $60°$, and becomes negligible when $\Theta = 30°$. This three-fold rotational symmetry can also be seen clearly from the $120°$ periodicity of the **H** canting angle $\theta_c$ shown in Fig. 3(d). $\theta_c$ is defined in Fig. 3(a) as the position of the $R_{2\omega}$ peak in the *zy* scans (green) with respect to $\theta = 90°$. The magnetic field in the *zy* scans is always perpendicular to the current and has both in-plane and out-of-plane components. Whereas the in-plane BMER has the same field angle dependence irrespective of $\Theta$, the $\Theta$ dependence of $\theta_c$ demonstrates the three-fold rotational symmetry of the out-of-plane component of the BMER. The out-of-plane spin component that underpins this signal must also have three-fold rotational symmetry. However, ARPES studies have reported a six-fold symmetric Fermi surface for the STO (111) surface 2DEG [13,14]. We conclude that the out-of-plane component of the spin texture in our BMER experiment breaks the symmetry of the electronic dispersion.

To investigate the origin of this signal, we have performed tight-binding supercell calculations for the STO (111) 2DEG based on a relativistic density functional theory calculation of the bulk band structure [32]. The supercell used for calculation has 90 Ti layers. Consistent with previous calculations and ARPES experiments [13,14], the Fermi surface for a 2DEG shown in Fig. 3(e) has a six-fold rotational symmetry. This reflects the equivalence of the $t_{2g}$ orbitals in the (111) plane and thus the orbitally-insensitive effect of quantum confinement on the bulk conduction band [13,14]. As shown in the inset of Fig. 3(e), in contrast to previous calculations, spin degeneracy is lifted due to the inclusion of atomic SOC in addition to an inversion symmetry breaking confinement potential. We find this Rashba spin-splitting to be less than 1 meV throughout the band structure, leading to very closely spaced Fermi surface sheets. Figure 3(f) shows the spin texture of the outermost subband; the behavior of higher order subbands is qualitatively similar. The strength and direction of the in-plane component of the spin expectation value is represented by the black arrows and has the form of a conventional Rashba field, locking spins perpendicularly to the momentum. It is consistent with the identical φ dependence of $R_{2\omega}$ in the *xy* scans for different Θ in Fig. 3(a-c).

In a free electron Rashba 2DEG system, the spin texture lies entirely in the plane of the surface or interface. In contrast, our calculation predicts a significant out-of-plane spin expectation value $\langle S_{111} \rangle$ for the subbands of the STO (111) 2DEG. $\langle S_{111} \rangle$ is represented by the red/white/blue color scale in Fig. 3(f) and shows a clear three-fold symmetry. In the presence of an out-of-plane magnetic field, this spin texture will result in a deformation of the Fermi surface with three-fold rotational symmetry. Therefore, $\langle S_{111} \rangle$ in our calculation supports the identification of out-of-plane spin as the origin of the three-fold rotationally symmetric BMER signal in the *zx* and *zy* scans of Figs. 3(a-c). The maximum canting angle of the spin vector with respect to the (111) plane for the largest Fermi surface sheet in our calculation is ~24° and along the Θ = 0° direction, which is comparable to the experimental canting angle we find in Fig. 3(d). The good agreement of the calculated spin texture with the symmetries of both the in-plane and out-of-plane BMER signal provides a compelling evidence for a Rashba-like spin splitting in the STO (111) 2DEG, with a three-fold symmetric out-of-plane spin component in addition to the conventional in-plane component with spin-momentum locking. The recent experimental and theoretical demonstration of the same characteristic BMER features for the $Bi_2Se_3$ TSS [26] further supports this conclusion, since the hexagonally warped Fermi surface of the TSS is qualitatively similar to a single Fermi surface sheet in our calculation of the STO (111) 2DEG.

Each of the six lobes of the Fermi surface in our calculation has a single dominant $t_{2g}$ orbital character. This reveals the preservation of the bulk cubic crystal field in the (111) 2DEG, which obeys $C_3$ symmetry. The out-of-plane spin component in the STO (111) 2DEG can be understood as another manifestation of the influence of the intrinsic crystal field on the subband structure. The preferential direction for the spin in the cubic crystal field of STO is along ⟨001⟩. Since the bulk crystal field is comparable to the band bending confinement potential, this preferential axis is partially preserved in the 2DEG. Therefore when, as for the STO (111) surface, the preferential spin polarization axis does not lie in the confinement plane, 2DEGs will naturally exhibit an out-of-plane spin component.

To characterize the magnitude of the BMER signal, we define the coefficient $\chi = 2\Delta R_{2\omega}/(R \cdot I \cdot H)$, which describes the BMER effect under unit electric and magnetic fields. As shown in Figs. 4(a) and 4(b), $\chi$ of our STO (111) 2DEGs shows a strong dependence on temperature and $n_H$ [32]. A further tuning of $\chi$ can be achieved by applying a back-gate voltage $V_g$, as shown in Fig. 4(c). $\chi$ can be modulated by one order of magnitude by gating, which we utilized to increase the signals for the experiments in Fig. 3. We find an $n_H^{-3}$ dependence of $\chi$ at 2 K. This dependence is predicted to arise in a single band 2D

Rashba system by the single relaxation time approximation of the Boltzmann equation [25]. However, the strong enhancement of $\chi$ that we observe below ~50 K cannot be explained within the same framework. Equally surprising is the large magnitude of $\chi$, which reaches 20 $A^{-1} T^{-1}$ at 2 K in our STO (111) 2DEG. This value is an order of magnitude larger than that found for a Rashba system whose spin-splitting of 75 meV is larger than the nominal bandwidth of the STO (111) 2DEG [25]. It also exceeds the value found for the TSS of $Bi_2Se_3$ which has only a single Fermi surface sheet [26]. This is remarkable because, within the single relaxation time approximation of the Boltzmann equation, the opposite helicities of the closely spaced Fermi surface sheets of the STO (111) 2DEG, as shown in Fig. 3(e), should lead to a small BMER signal. Our data thus question the validity of the single relaxation time approximation in STO (111) 2DEGs and highlights the possibility of a large effect arising from subband dependent scattering rates. The large $\chi$ in our samples despite only a very small spin-splitting further suggests advances in the theory of second order conductivity in oxides, for example considering coherent superposition of spin states [38] or magnetic breakdown between the very closely spaced Fermi surface sheets [39], are needed to achieve a quantitative understanding.

In summary, we have revealed the 3D nature of the spin texture of the STO (111) 2DEG by probing the BMER. Supported by band structure calculations, we conclude that the STO (111) 2DEG has a spin-split subband structure with a three-fold symmetric out-of-plane spin component in addition to the conventional spin-momentum locked in-plane component of the free electron Rashba model. Broken inversion symmetry along the surface normal results in the in-plane Rashba field while the out-of-plane component is a manifestation of the strong cubic crystal field of bulk STO. Thus, our calculations demonstrate that the origin of this spin texture can be traced back to the choice of quantum confinement direction. More broadly our results suggest that lattice induced spin canting is a generic property of oxide 2DEGs. The BMER response in STO (111) is unexpectedly high suggesting that the physics of this system and of the BMER effect are not yet fully understood on a quantitative level, and also that STO (111) has potential for spintronics in spite of the small magnitude of the spin splitting predicted by our calculations. Moreover, our study opens pathways for the manipulation of current-induced spin polarizations using different current directions, which is appealing for spintronic applications. In addition, the distinctive tunability by electric gating offers a high flexibility for engineering spin-dependent properties. Finally, our work also establishes BMER as a sensitive probe of the spin texture in oxide 2DEGs.

We thank Y. Z. Luo, J. Zhou, W. X. Zhou, K. Han, and S. W. Zeng for useful discussions. This work was partially supported by the National Research Foundation (NRF), Prime Minister's Office, Singapore, under its Competitive Research Programme (CRP Award No. NRFCRP12-2013-01). F.B. and S.M.W. were supported by SNSF 200020-165791. F.Y.B. was supported by SNSF Ambizione PZ00P2-161327. M.S.B. gratefully acknowledges support from the CREST, JST (JPMJCR16F1). Work by S.S.-L.Z. and O.H. were supported by the Department of Energy, Office of Science, Basic Energy Sciences, Materials Sciences and Engineering Division. G.V. gratefully acknowledges support for this work from NSF Grant DMR-1406568.

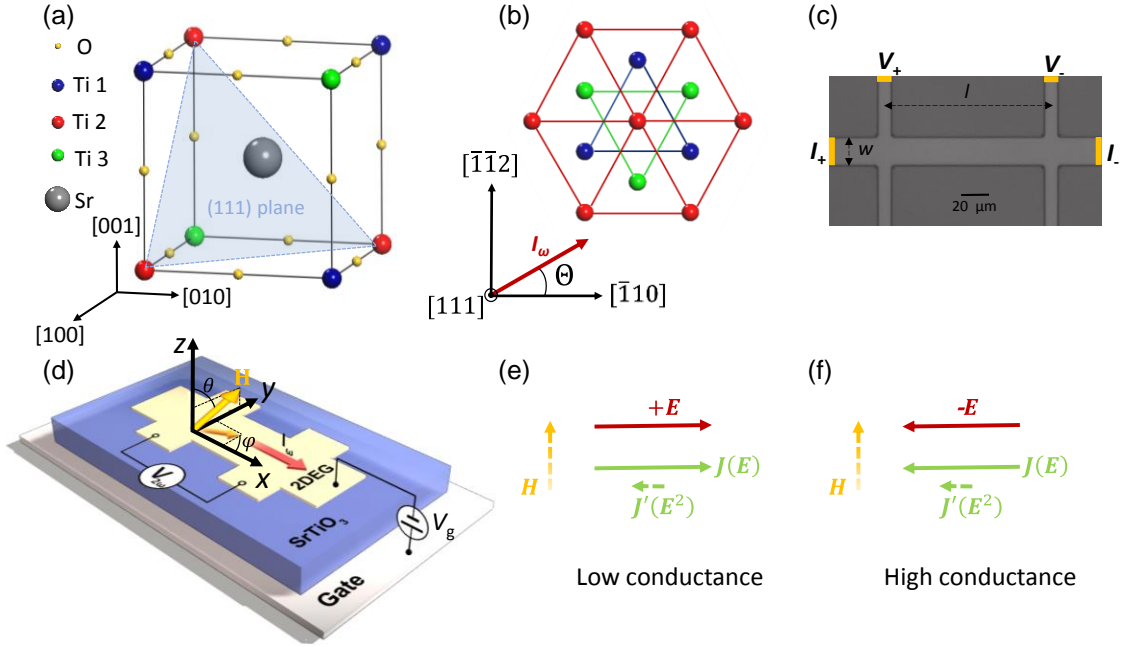

Fig. 1. (a) Schematic unit cell of the cubic perovskite crystal STO. The (111) plane is indicated in shade. (b) Top view of three consecutive Ti layers. A bilayer forms a honeycomb lattice. The Hall bar angle Θ between the current direction and the $[\bar{1}10]$ axis is indicated. (c) Optical image of a Hall bar. *l* and *w* indicate the length and width of current channel. (d) Schematic for the second harmonic magnetoresistance measurements on the 2DEG of STO (111) surface. A sinusoidal current $I_\omega$ was applied, and the second harmonic longitudinal voltage $V_{2\omega}$ was measured under a magnetic field **H** in either *xy*, *zy* or *zx* plane. (e), (f) For **H** applied perpendicular to the in-plane electric field **E**, a nonlinear charge current $\mathbf{J}'(\mathbf{E}^2)$ is generated in the 2DEG at the STO (111) surface in addition to the linear current $\mathbf{J}(\mathbf{E})$, which gives rise to a low conduction state under +**E** (e) and a high conduction state under –**E** (f).

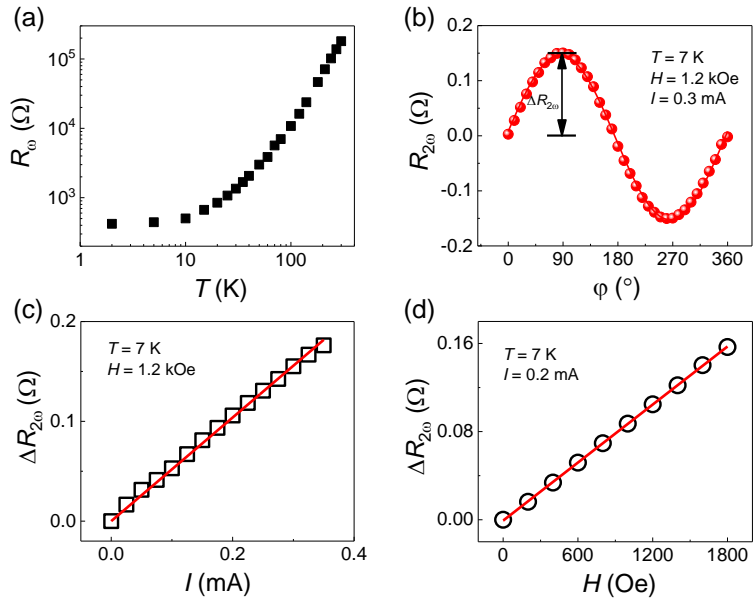

Fig. 2. (a) Temperature dependence of the first order device resistance $R_\omega$. (b) Angular dependence of $R_{2\omega}$ while rotating **H** in the *xy* plane. A vertical offset was subtracted for clarity. The solid line is a $\sin(\varphi)$ fit to the data. A linear dependence of $\Delta R_{2\omega}$ on the current *I* (c) and magnetic field *H* (d). The solid lines are linear fits to the data from sample 1.

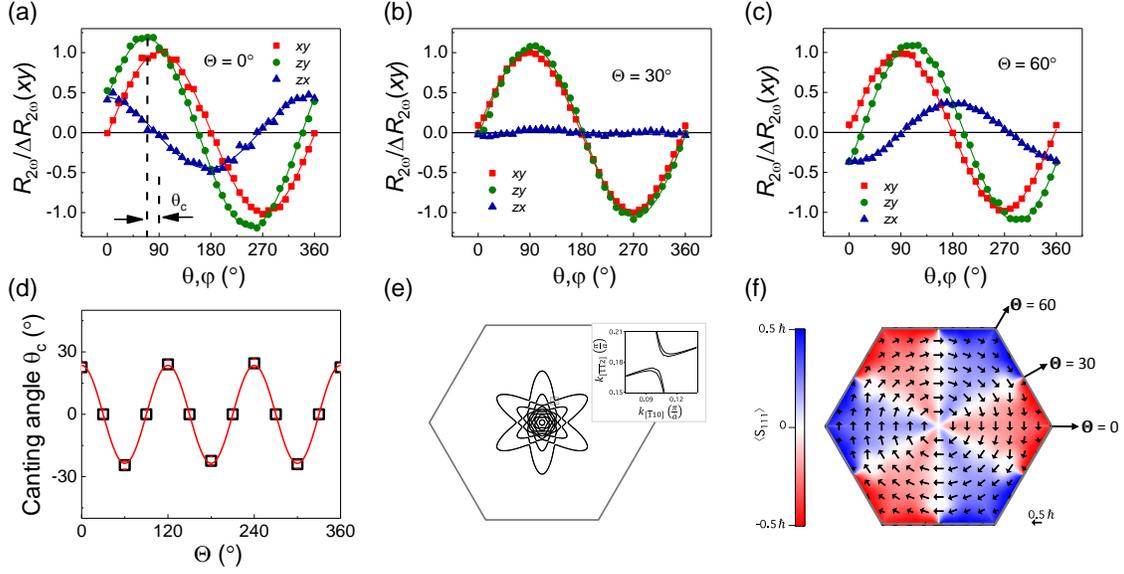

Fig. 3. $R_{2\omega}$ (normalized by $\Delta R_{2\omega}$ of the *xy* scan) for scans of **H** in the *xy* (red), *zy* (green) and *zx* (blue) planes with the Hall bar angle $\Theta = 0°$ (a), 30° (b) and 60° (c). The solid lines are fits to $\sin\varphi(\theta)$. (d) The magnetic field canting angle $\theta_c$, defined in (a) as the angle position of $R_{2\omega}$ peak in the *zy* scan with respect to $\theta = 90$, as a function of $\Theta$. Measurements were performed under $H = 5$ kOe, $I = 0.5$ mA, $V_g = -60$ V and $T = 2$ K for sample 5. (e) The Fermi surface of a self-consistent tight binding supercell calculation for a 2DEG at the STO (111) surface with carrier density $2.1 \times 10^{14}$ cm$^{-2}$. The inset is zoom of shaded area. (f) The spin texture of the highest density Fermi surface sheet. The in-plane spin component $\sqrt{\langle S_{\overline{1}\overline{1}2}\rangle^2 + \langle S_{\overline{1}10}\rangle^2}$ is represented by black arrows and the out-of-plane spin component $\langle S_{111}\rangle$ is represented by the red/white/blue color scale.

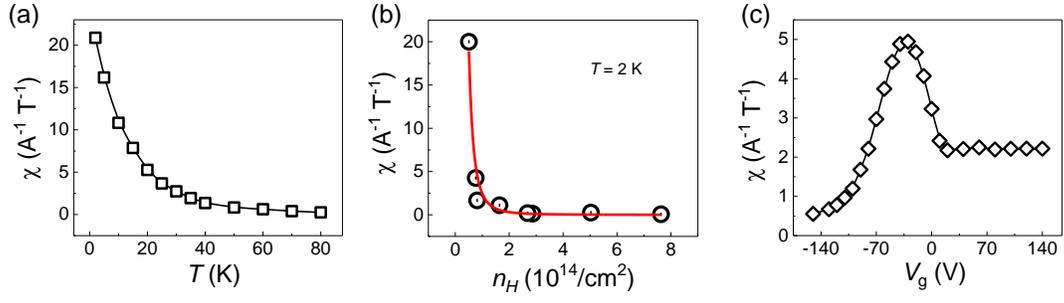

Fig. 4. (a) Temperature dependence of $\chi$ for sample 1 in the *xy* scan. (b) $\chi$ as a function of the carrier density $n_H$ extracted from various samples with different Ar$^+$ irradiation treatments at 2 K. The red line is a fit to $n_H^{-3}$. (c) Gate voltage dependence of $\chi$ under $H = $ 5 kOe, $I = 0.5$ mA and $T = 2$ K in the *zy* scan for sample 5.